\documentclass[iop,revtex4]{emulateapj}
\pdfoutput=1

\newcommand{\degree}{^\circ}

\usepackage{amsmath}
\usepackage{rotating}

\shorttitle{AB Aur Spiral Movement}
\shortauthors{Lomax et al.}

\begin{document}

\title{Constraining the Movement of the Spiral Features and the Locations of Planetary Bodies within the AB Aur System}

\author{Jamie R. Lomax\altaffilmark{1},
John P. Wisniewski\altaffilmark{1},
Carol A. Grady\altaffilmark{2,3,4},
Michael W. McElwain\altaffilmark{35},
Jun Hashimoto\altaffilmark{1},
Tomoyuki Kudo\altaffilmark{5},
Nobuhiko Kusakabe\altaffilmark{6},
Yoshiko K. Okamoto\altaffilmark{40},
Misato Fukagawa\altaffilmark{28},
Lyu Abe\altaffilmark{7},
Wolfgang Brandner\altaffilmark{8},
Timothy D. Brandt\altaffilmark{9},
Joseph C. Carson\altaffilmark{10},
Thayne M Currie\altaffilmark{5},
Sebastian Egner\altaffilmark{5},
Markus Feldt\altaffilmark{8},
Miwa Goto\altaffilmark{11},
Olivier Guyon\altaffilmark{5},
Yutaka Hayano\altaffilmark{5},
Masahiko Hayashi\altaffilmark{6},
Saeko S. Hayashi\altaffilmark{5},
Thomas Henning\altaffilmark{8},
Klaus W. Hodapp\altaffilmark{12},
Akio Inoue\altaffilmark{28},
Miki Ishii\altaffilmark{6},
Masanori Iye\altaffilmark{6},
Markus Janson\altaffilmark{37},
Ryo Kandori\altaffilmark{6},
Gillian R. Knapp\altaffilmark{13},
Masayuki Kuzuhara\altaffilmark{14},
Jungmi Kwon\altaffilmark{15},
Taro Matsuo\altaffilmark{16},
Satoshi Mayama\altaffilmark{17},
Shoken Miyama\altaffilmark{18},
Munetake Momose\altaffilmark{30},
Jun-Ichi Morino\altaffilmark{6},
Amaya Moro-Martin\altaffilmark{38,39},
Tetsuo Nishimura\altaffilmark{5},
Tae-Soo Pyo\altaffilmark{5},
Glenn H Schneider\altaffilmark{31},
Eugene Serabyn\altaffilmark{20},
Michael L. Sitko\altaffilmark{32,33},
Takuya Suenaga\altaffilmark{6,21},
Hiroshi Suto\altaffilmark{6},
Ryuji Suzuki\altaffilmark{6},
Yasuhiro H. Takahashi\altaffilmark{6,15},
Michihiro Takami\altaffilmark{22},
Naruhisa Takato\altaffilmark{5},
Hiroshi Terada\altaffilmark{5},
Christian Thalmann\altaffilmark{23},
Daigo Tomono\altaffilmark{5},
Edwin L. Turner\altaffilmark{9},
Makoto Watanabe\altaffilmark{25},
Toru Yamada\altaffilmark{26},
Hideki Takami\altaffilmark{6},
Tomonori Usuda\altaffilmark{15},
Motohide Tamura\altaffilmark{6,15}}
\altaffiltext{1}{Homer L. Dodge Department of Physics, University of Oklahoma, Norman, OK 73071, USA; Jamie.R.Lomax@ou.edu, wisniewski@ou.edu}
\altaffiltext{2}{Exoplanets and Stellar Astrophysics Laboratory, Code 667, Goddard Space Flight Center, Greenbelt, MD
20771, USA; carol.a.grady@nasa. gov}
\altaffiltext{3}{Eureka Scientific, 2452 Delmer, Suite 100, Oakland CA 96002, USA}
\altaffiltext{4}{Goddard Center for Astrobiology}
\altaffiltext{5}{Subaru Telescope, National Astronomical Observatory of Japan, 650 North A'ohoku Place, Hilo, HI 96720, USA}
\altaffiltext{6}{National Astronomical Observatory of Japan, 2-21-1, Osawa, Mitaka, Tokyo, 181-8588, Japan}
\altaffiltext{7}{Laboratoire Lagrange (UMR 7293), Universite de Nice-Sophia Antipolis, CNRS, Observatoire de la Cote d'Azur, 28 avenue Valrose, F-06108 Nice Cedex 2, France}
\altaffiltext{8}{Max Planck Institute for Astronomy, K\"{o}nigstuhl 17, D-69117 Heidelberg, Germany}
\altaffiltext{9}{Astrophysics Department, Institute for Advanced Study, Princeton, NJ 08540, USA}
\altaffiltext{10}{Department of Physics and Astronomy, College of Charleston, 58 Coming St., Charleston, SC 29424, USA}
\altaffiltext{11}{Universit\"{a}ts-Sternwarte M\"{u}nchen, Ludwig-Maximilians-Universit\"{a}t, Scheinerstr. 1, D-81679 M\"{u}nchen, Germany}
\altaffiltext{12}{Institute for Astronomy, University of Hawaii, 640 N. A'ohoku Place, Hilo, HI 96720, USA}
\altaffiltext{13}{Department of Astrophysical Science, Princeton University, Peyton Hall, Ivy Lane, Princeton, NJ 08544, USA}
\altaffiltext{14}{Department of Earth and Planetary Sciences, Tokyo Institute of Technology, 2-12-1 Ookayama, Meguro-ku,
Tokyo 152-8551, Japan}
\altaffiltext{15}{Department of Astronomy, The University of Tokyo, 7-3-1, Hongo, Bunkyo-ku, Tokyo, 113-0033, Japan}
\altaffiltext{16}{Department of Astronomy, Kyoto University, Kitashirakawa-Oiwake-cho, Sakyo-ku, Kyoto 606-8502, Japan}
\altaffiltext{17}{The Center for the Promotion of Integrated Sciences, The Graduate University for Advanced Studies (SOKENDAI), Shonan International Village, Hayama-cho, Miura-gun, Kanagawa 240-0193, Japan}
%\altaffiltext{}{Department of Earth and Planetary Sciences, Tokyo Institute of Technology, Ookayama, Meguro-ku, Tokyo 152-8551, Japan}
%\altaffiltext{12}{Department of Astronomy \& Astrophysics, University of Toronto, 50 George St., Toronto, Ontario, M5S 3H4, Canada}
\altaffiltext{18}{Hiroshima University, 1-3-2, Kagamiyama, Higashi-Hiroshima 739-8511, Japan}
\altaffiltext{19}{Department of Astrophysics, Departamento de Astrofisica, CAB (INTA-CSIC), Instituto Nacional de T\'{e}cnica Aeroespacial, 28850 Torrej\'{o}n de Ardoz, E-28850 Madrid, Spain}
\altaffiltext{20}{Jet Propulsion Laboratory, California Institute of Technology, Pasadena, CA, 91109, USA}
\altaffiltext{21}{Department of Astronomical Science, The Graduate University for Advanced Studies, 2-21-1, Osawa, Mitaka, Tokyo, 181-8588, Japan}
\altaffiltext{22}{Institute of Astronomy and Astrophysics, Academia Sinica, P.O. Box 23-141, Taipei 10617, Taiwan}
\altaffiltext{23}{Institute for Astronomy, ETH Zurich, Wolfgang-Pauli-Strasse 27, 8093 Zurich, Switzerland}
\altaffiltext{24}{Kavli Institute for Physics and Mathematics of the Universe, The University of Tokyo, 5-1-5, Kashiwanoha, Kashiwa, Chiba 277-8568, Japan}
\altaffiltext{25}{Department of Cosmosciences, Hokkaido University, Kita-ku, Sapporo, Hokkaido 060-0810, Japan}
\altaffiltext{26}{Astronomical Institute, Tohoku University, Aoba-ku, Sendai, Miyagi 980-8578, Japan}
\altaffiltext{28}{Graduate School of Science, Osaka University, 1-1 Machikaneyama, Toyonaka, Osaka 560-0043, Japan}
\altaffiltext{30}{College of Science, Ibaraki University, Bunkyo 2-1-1, Mito, 310-8512 Ibaraki, Japan}
\altaffiltext{31}{Steward Observatory, 933 N. Cherry Ave., University of Arizona, Tucson, Arizona 85721, USA}
\altaffiltext{32}{Department of Physics, University of Cincinnati, Cincinnati, OH 45221, USA}
\altaffiltext{33}{Space Science Institute, 475 Walnut Street, Suite 205, Boulder, CO 80301, USA}
\altaffiltext{35}{NASA Goddard Space Flight Center, Code 6681, Greenbelt, MD 20771, USA}
\altaffiltext{37}{Department of Astronomy, Stockholm University, AlbaNova University Center, SE-10691 Stockholm, Sweden}
\altaffiltext{38}{Space Telescope Science Institute, 3700 San Martin Dr., Baltimore, MD 21218, USA}
\altaffiltext{39}{Center for Astrophysical Sciences, Johns Hopkins University, Baltimore, MD 21218, USA}
\altaffiltext{40}{Institute of Astrophysics and Planetary Sciences, Faculty of Science, Ibaraki University, 2-1-1 Bunkyo, Mito, Ibaraki 310-8512, Japan}

\begin{abstract}
We present new analysis of multi-epoch, \textit{H}-band, scattered light images of the AB Aur system. We used a Monte Carlo, radiative transfer code to simultaneously model the system's SED and \textit{H}-band polarized intensity imagery. We find that a disk-dominated model, as opposed to one that is envelope dominated, can plausibly reproduce AB Aur's SED and near-IR imagery. This is consistent with previous modeling attempts presented in the literature and supports the idea that at least a subset of AB Aur's spirals originate within the disk. In light of this, we also analyzed the movement of spiral structures in multi-epoch \textit{H}-band total light and polarized intensity imagery of the disk. We detect no significant rotation or change in spatial location of the spiral structures in these data, which span a 5.8 year baseline. If such structures are caused by disk-planet interactions, the lack of observed rotation constrains the location of the orbit of planetary perturbers to be $>$47 AU.
\end{abstract}

\keywords{}

\section{Introduction} 

AB Aur (also known as HD 31293 and SAO 57506, $d=144$ pc) is a young, ($4\pm 1$ Myr) intermediate-mass ($2.4\pm 0.2$ M$_\sun$), Herbig Ae star \citep{van den Ancker,DeWarf2} that is actively accreting material. It is surrounded by a large envelope which extends out to at least 1320 AU and blends into a nearby nebula \citep{Grady}. Within the envelope, a protoplanetary disk (r $\sim 450$ AU; Mannings \& Sargent 1997) surrounds the central star and displays many complex structures \citep{Hashimoto,Fukagawa,Grady}.  

At 1.6 and 2 $\mu$m the disk has a region of decreased polarized intensity, which is likely due to the scattering geometry of the surface of AB Aur's inclined disk \citep{Oppenheimer,Perrin}. However, more recent HiCIAO \textit{H}-band imagery has shown an additional six regions of decreased polarized intensity (PI) that are not explained by geometric scattering effects \citep{Hashimoto}. These data also revealed an approximately 16 AU wide gap in the disk. Centered at 80 AU from the central star, it is similar to the mid-IR gap inferred by \cite{Honda} from models of their 24.6 $\mu$m imagery, but appears to be different from the gap detected by \cite{Tang}, who find a gap which ends at approximately 110 AU and is about 90 AU wide in 1.3 mm continuum emission.

Spiral structures in the disk were first detected in STIS imagery by \cite{Grady} and later imaged in the \textit{H} band by \cite{Fukagawa}, who suggested that they are either maintained by a planet, due to gravitational instabilities within the disk, or the result of the outer envelope replenishing disk material. Subsequent work by \cite{Hashimoto} and \cite{Lin} show evidence of the spiral structures in \textit{H}-band PI imagery, $^{12}$CO (3-2) maps, and at 850 $\mu$m. Both works favor planetary bodies perturbing the disk as an explanation for the formation of the spirals. However, a planet has yet to be detected in the system.

\cite{Tang} recently detected four spirals in the CO gas, which are generally not coincident with the spirals detected in the near-IR. For example, \cite{Tang}'s CO S2 spiral appears to share a base with the \textit{H}-band S1 spiral as labeled by \cite{Hashimoto}, but the outer regions of the two spiral arms do not overlap. Similarly, the \textit{H}-band S3 spiral appears to potentially be a continuation of the CO S3 spiral at regions farther from the central star, but there is a gap between the regions where the spirals have been detected; the innermost region detected of the \textit{H}-band S3 spiral and the outermost detected region for the CO S3 spiral do not spatially overlap. Finally, the other two CO spirals, CO S1 and CO S4, appear to have no near-IR counterpart at all.

\cite{Tang} recently suggested a different formation mechanism for the spirals whereby a combination of the rotation and infall of material from the envelope allows for the build up of higher density regions along the envelope's bipolar cavities. Due to the system's low inclination angle ($i=22\degree$; Tang et al. 2012), these regions of higher density are projected onto the disk and form the observed spiral structures. They appear as part of the disk, but are actually high density regions of the envelope. However, observational constraints on the density, infall rate, and rotational speed of the envelope do not exist, making it difficult to determine the likelihood of this scenario.

Very little is known about AB Aur's envelope, partly because of the difficulty of disentangling the observed contributions of the disk and envelope (e.g. CO lines trace the mid-plane disk structure, but \citealt{Tang} also suggests it traces the envelope morphology). \cite{Pietu} found no evidence for any infall of material in their study of the CO lines. This agrees with the overall conclusions of \cite{Robitaille}, who use two dimensional radiative transfer modeling of the system's SED to place constraints on the mass accretion rate from the envelope. They find that the infall rate might be as high as $10^{-6}$ M$_\sun$ yr$^{-1}$, but their best fit model uses no infall at all, suggesting that the envelope is very optically thin.

In this paper, we analyze multi-epoch \textit{H}-band imagery of AB Aur to investigate whether the positions of its spiral arms at these wavelengths have changed with time. A variety of observations of AB Aur have been modeled in the past (including but not limited to, its SED by \citealt{Bouwman} and \citealt{Robitaille}, SED and NIR interferometry by \citealt{tan08}, NIR scattered light imagery by \citealt{Perrin} and \citealt{jan10}, SED and mid-IR imagery by \citealt{Honda}, and mm emission by \citealt{Pietu}). However, self-consistent models of the SED and near-IR imagery of the system, which can be useful in interpreting multi-epoch imagery, have not been extensively explored. Therefore, we first used a three-dimensional, Monte Carlo, radiative transfer code to model the overall behavior of the system's SED and \textit{H}-band imagery. After finding that some of the spiral structures in the system could arise in the disk, as noted in previous works, we compare two sets of archival \textit{H}-band imagery in order to determine if the position of the spirals have changed with time.

\section{Observations} 

\subsection{Spectral Energy Distribution}
We used the SED compiled by \cite{Robitaille} in their Tables 8, 9, 10, and 11 to compare to our modeled SED. This includes \textit{UBVRI} and \textit{LMNQ} data from Kenyon \& Hartmann (1995); \textit{JHK} data from the 2MASS all-sky survey; far-IR \textit{IRAS} 12 $\mu$m, 25 $\mu$m, 60 $\mu$m, and 100 $\mu$m data from Weaver \& Jones (1992); and SHARC 350 $\mu$m, SCUBA 450 $\mu$m, and SCUBA 850 $\mu$m submillimeter data from Andrews \& Williams (2005).  We supplemented the \cite{Robitaille} SED with additional photometry from the AllWISE Catalog at 3.35, 11.6, and 22.1 $\mu$m \citep{Cutri}; the Akari IRC All-Sky Survey Point Source Catalogue data at 8.61 and 18.4 $\mu$m; Hershall data at 70 and 160 $\mu$m \citep{Pascual}; SCUBA-2 1300 $\mu$m data \citep{Mohanty}; and SMA data at 1300 $\mu$m \citep{Andrews}. We also made use of spectra from the Short Wavelength Spectrometer (SWS) aboard the \textit{Infrared Space Observatory} (ISO) \citep{van den Ancker2000}.

\subsection{\textit{H} Band Imagery}

We also use two archival \textit{H}-band images of AB Aur obtained by the CIAO \citep{Tamura1998} and HiCIAO \citep{Tamura} instruments on the Subaru 8.2 m Telescope. The CIAO data (originally published in Fukagawa et al. 2004) consist of several image data sets taken on 2004 January 8 and 11 which were combined into the one final \textit{H}-band image with a pixel scale of $21.33 \pm 0.02$ mas pixel$^{-1}$. Additional details about how those observations were obtained, reduced, and calibrated can be found in \cite{Fukagawa}, including information about exposure times, number of frames, and PSF subtraction.

The HiCIAO data (originally published in Hashimoto et al. 2011) were obtained on 2009 October 31 as part of the Strategic Explorations of Exoplanets and Disks with Subaru (SEEDS) program \citep{Tamura2009}. Seven sets of data were combined to form the final \textit{H}-band PI image ($9.3 \pm 0.02$ mas pixel$^{-1}$). Additional details about observing in polarimetric mode and reduction procedures can be found in \cite{Hashimoto}.

\section{\texttt{HOCHUNK3D} Models of AB Aur}

AB Aur has a rich history both of multi-wavelength observations and detailed modeling efforts to explore the nature of the gas and dust surrounding the system, as noted in the Introduction.  As our goal is to analyze and interpret multi-epoch \textit{H}-band imagery of the system, we first modeled the global behavior of the system's SED and \textit{H}-band imagery to assess the potiential origin of observed morphological structures.  We used \texttt{HOCHUNK3D}, a publicly available Monte Carlo radiative transfer code (MCRT) that allows a user to place circumstellar material around a forming star and define its three-dimensional geometry (a description of the original code can be found in Whitney et al. 2003; see Whitney et al. 2013 and references therein for updates to the code) to simultaneously model SEDs and imagery. Currently, the code offers a suite of different geometries, including warps, gaps, and spirals within the accretion disk structure; an infalling envelope; and a bipolar outflow cavity. In the most recent update to the code, \cite{Whitney2013} decouples the large and small grain populations so that disk settling can be included. 

The analytic formula for calculating these geometries are given in \cite{Whitney2003}; however, we briefly describe them and some additional details here. The distribution of dust in disks surrounding a central star is controlled by the $\alpha$ and $\beta$ parameters in the following density profile equations: 
\begin{equation}
\label{eqn:density}
\rho \propto r^{- \alpha} \exp\left\{ \left[ \frac{z}{H}\right]^2\right\}
\end{equation}
\begin{equation}
\label{eqn:scale_height}
H \propto r^{-\beta}
\end{equation}
where H is the scale height of the disk, r is the radius, and z is the distance above and below the midplane of the disk. Accretion from the disk onto the central star is included by taking into account the accretion luminosity of the system and is set by an accretion rate parameter. The functional form of the envelope in our model is given by \cite{Ulrich} and is a rotating sphere undergoing free fall gravitational collapse. Temperatures for the various components are corrected based on the Lucy method \citep{Lucy}. The code uses a Henyey-Greenstein phase function for scattering photons in this material, and calculates output SEDs and imagery for a given viewing angle using the `peeling-off' optical raytracing algorithm \citep{Whitney2013}.

We modeled the AB Aur system as a pre-transitional disk with the \texttt{HOCHUNK3D} code. Our basic disk geometry includes a settled, large grain disk surrounded by smaller grains, and a gap at 100 AU (consistent with the location Hashimoto et al. 2011 derived from \textit{H}-band PI observations). We use a 10,000 K Kurucz model atmosphere file with $log(g)=3.5$ and $log\frac{Z}{H}=-1.5$ as our input stellar spectrum and determined basic disk parameters through a trial and error process. While the \texttt{HOCHUNK3D} code has the ability to add spirals into a disk structure, it is done in an ad hoc fashion and requires us to make an a priori decision about their location, i.e. determining that the spirals are disk or envelope structures. Therefore, we do not include any non-axisymmetric structures within our model, and rather focus on reproducing the global SED and overall surface brightness morphology of the system.

We computed $\sim$500 unique models that explored a range of disk and envelope parameter space.  We began this process using 1,000,000 photon model runs to compare the observed SED to those produced by our model. We formally calculated a $\chi^2$ value for each model in our parameter space and compared the $\chi^2$ trends of different parameters (e.g. different envelope sizes) against each other. However, we caution that formal goodness of fit metrics such as $\chi^2$ are generally not useful for determining which \texttt{HOCHUNK3D} models better fit observed data because it is well known that MCRT modeling involves significant parameter degeneracies (see e.g. \citealt{Robitaille}). It is highly likely that a `better' model, as indicated by the $\chi^2$ statistic, can be found using unphysical parameters.  Hence, we did not blindly follow the $\chi^2$ statistic to determine which model best reproduces the data.  Instead, we use this formal statistic only to compare trends in our models which already have parameters appropriate for the AB Aur system.  Once we arrived at parameter families that broadly reproduced the observed SED, we followed those with $\sim$270 runs with 100,000,000 photons to compute detailed model imagery of the system in the H-band. 

Because the broad parameter space using a grid of slightly simpler, albeit generally analogous, models \citep{rob06} was analyzed for AB Aur and constrained by its observed SED \citep{Robitaille}, and because the goal of our modeling was simply to establish a plausible `best-fit' to the system's SED and H-band imagery, we do not describe the detailed properties of the acceptable model parameter families we found.  Rather, we simply present and discuss the basic properties of this `best-fit' model (Table 1), with the caveat that it is not unique owing to well-known parameter degeneracies, and use this as a basis to interpret the available multi-epoch imagery of the system.

\begin{deluxetable}{lccc}
\tablecolumns{4}
\tablewidth{0pt}
%\tabletypesize{\scriptsize}
\tablecaption{AB Aur Star and Disk Parameters Used in \texttt{HOCHUNK3D} Models \label{DParams}}
\startdata
Parameter & Assumed Value& Reference\\
\hline
\multicolumn{3}{l}{\textbf{Central Source Properties}} \\
\hline
    M$_\star$ (M$_\sun$) & 2.4 & 1, 2\\
    R$_\star$ (R$_\sun$) & 2.5 & 3\\
    Teff (K) & 9520 & 3\\
    \hline
\multicolumn{3}{l}{\textbf{Overall Disk Properties}} \\
\hline
    M$_{\text{Disk}}$\tablenotemark{a} (M$_\sun$) & 0.032 & 4 \\
    R$_{\text{max}}$ (AU) & 420 & 5\\
    Fraction of Mass in the Settled Disk & 0.075 \\
    Gap Inner Radius (AU) & 0.55 \\
    Gap Outer Radius (AU) & 100 & 6\\
    %$\alpha_{\text{gap}}$ & 2 \\
    $\rho_{\text{gap}}$\tablenotemark{b} & 0.4 \\
    $i$ (degrees) & 20 & 7\\
          \hline
\multicolumn{3}{l}{\textbf{Grain Properties}} \\          
\multicolumn{3}{l}{\textit{Large Grain Disk}} \\          
\hline
    Thermal Dust Model & www003 & 8 \\
    $<$200 \AA Grains & draine & 9, 10 \\
    Fraction of mass $<$200 \AA grains & 0.2 \\
    Scale Height\tablenotemark{c} & 0.6 \\
    $\alpha$ & 2 \\
    $\beta$ & 1.35 \\
          \hline
\multicolumn{3}{l}{\textit{Small Grain Disk}} \\        
\hline
    Thermal Dust Model & kmh & 11 \\
    $<$200 \AA Grains & draine & 9, 10 \\
    Fraction of mass $<$200 \AA grains & 0.1 \\
    Scale Height\tablenotemark{c} & 1 \\
    $\alpha$ & 2 \\
    $\beta$ & 1.18 \\
\hline
    \multicolumn{3}{l}{\textbf{Envelope Properties}} \\
\hline
    Envelope Properties &  \\
    Thermal Dust Model & ice095\tablenotemark{a} \\
    $<$200 \AA Grains & draine\tablenotemark{b} \\
    Infall Rates (M$_\sun$ yr$^{-1}$) & 10$^{-7}$\\ 
    R$_{\text{max}}$ (AU) & 1320 \\
 \hline
\multicolumn{3}{l}{\textbf{Bipolar Cavity}} \\
\hline
    Thermal Dust Model & kmh\tablenotemark{c} \\
    $<$200 \AA Grains & draine\tablenotemark{b} \\
    Opening Angles ($\theta_1$; degrees)  & 20
\enddata
\tablecomments{Final adopted disk parameters determined from basic fitting (Section 4.1.1). References: (1) van den Ancher 1997; (2) DeWarf et al. 2003; (3) van den Ancher et al. 2002; (4) Robitaille et al. 2007; (5) Mannings \& Sargent 1997; (6) Hashimoto et al. 2011; (7) Blake \& Boogert 2004; (8) model 1 in Wood et al. 2002; (9) Draine \& Li 2007; (10) Wood et al. 2008; (11) average galactic ISM grains by Kim et al. 1994
\tablenotetext{a}{Total disk mass}
\tablenotetext{b}{This is the ratio of the density of the gap just inside the outer gap radius to just outside the outer gap radius.}
\textbf{\tablenotetext{c}{Scale heights are in units of the dust destruction radius.}}}
\end{deluxetable}

\begin{center}
\begin{figure*}[]
%\figurenum{4}
%\epsscale{1}
\centering
\includegraphics[width=\textwidth]{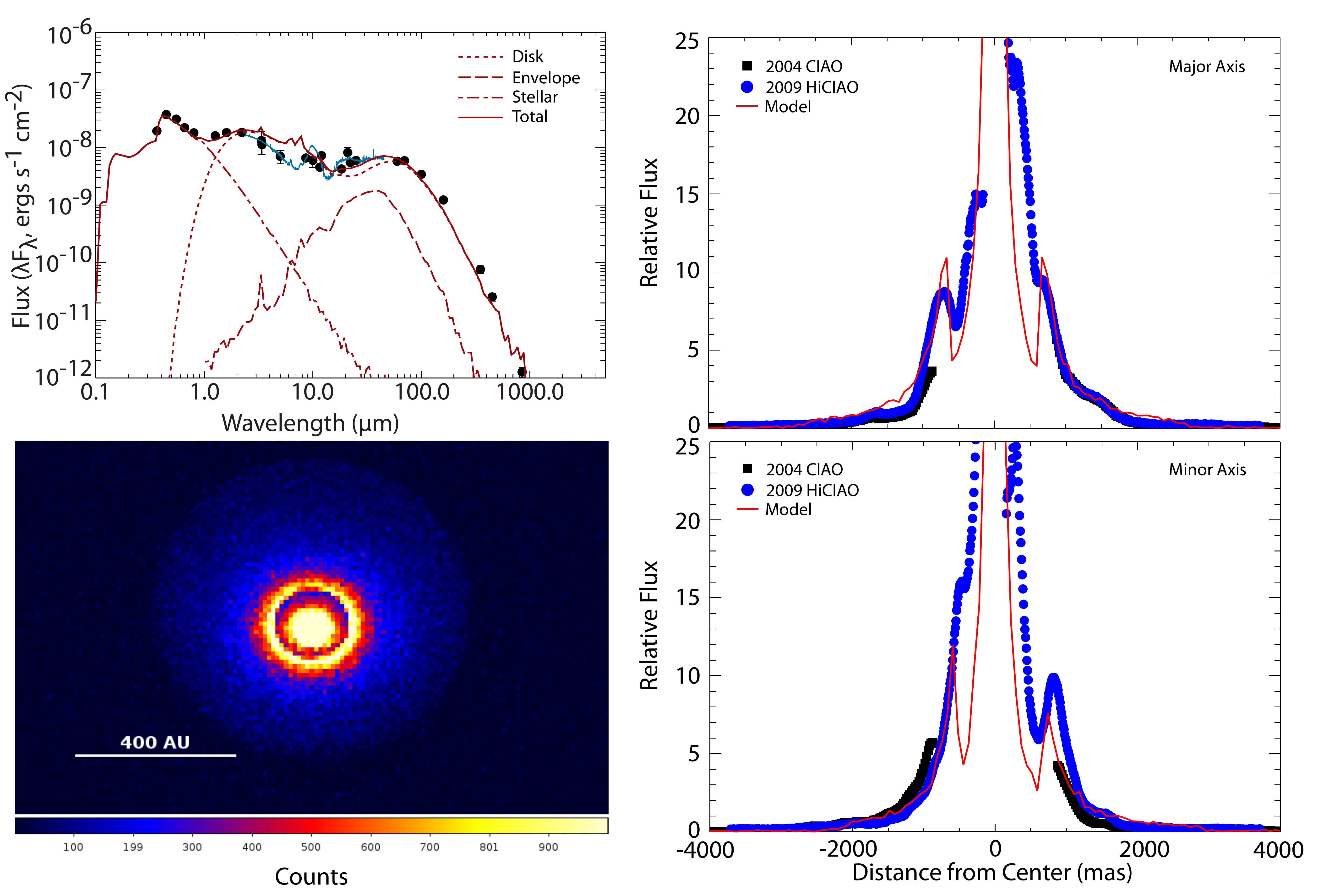}%
\caption{\textit{Left top:} Comparision of our model's SED (red solid curve) to the observed AB Aur SED (black filled circles and blue spectrum, see Section 2.1). The individual contributions from the envelope (red dashes), disk (red dots), and stellar (red dot dashed) components to our modeled SED are shown. Error bars are shown for uncertainties that are larger than the point size of our photometry. \textit{Left bottom:} \textit{H} band modeled PI imagery of the AB Aur system. \textit{Right:} Radial profile cross cuts of our model (solid red lines) compared to our archival imagery. The blue circles represent the HiCIAO data taken in 2009, while the black squares represent the CIAO data from 2004. The top panel displays the radial profiles across the major axis (position angle of 36$\degree$). The bottom panel displays the same information for the disk gap's minor axis. The flux scale is relative. We normalized all of the data to each other for display purposes. Data from regions of the images significantly affected by PSF-subtraction residuals and inside of the HiCIAO and CIAO inner working angles are not shown.} \label{SED}
\end{figure*}
\end{center}

The SED and \textit{H}-band PI imagery for our best-fit model are shown in Figure \ref{SED}. This model broadly reproduces AB Aur's SED, although we note it slightly over-predicts the level of near-IR flux compared to what is observed. Our model imagery (Figure \ref{SED}) reproduces the visibility of the gap within the disk.  Moreover, our model reproduces the general behavior of the observed surface brightness along the major and minor axes outside of the gap region in the archival \textit{H}-band PI imagery taken with HiCIAO (Figure \ref{SED}). Because previous maps of the polarization fraction of the disk are relatively uniform (i.e. \citealt{Perrin}) due to the system's low inclination, we also compare the surface brightness profiles of the \textit{H}-band total light CIAO imagery along the major and minor axes with our \textit{H}-band PI model imagery in Figure 1. With this comparison we find results similar to the comparison between our model and the HiCIAO data; our model reproduces the general behavior of the disk's observed surface brightness profile. However, our the model is slightly less successful at reproducing the observed surface brightness at the edge-of and interior to the gap region in the HiCIAO data. Because this flux discrepancy at and inside the disk edge can be reduced by increasing the scale height and density of the material in the gap region at the expense of degrading the NIR SED fit, we speculate that a broader exploration of geometries at this disk boundary, which are outside the scope of this paper, could improve these fits. We note that in order to compare our model with the observed data, we must normalize all of the surface brightness profiles relative to each other. This type of scaling is not uncommon when using HOCHUNK3D outputs because the model imagery is in units of counts as opposed to an absolute flux. We also scaled the CIAO and HiCIAO imagery relative to each other for display purposes.

Despite simultaneously modeling both the SED and \textit{H}-band PI imagery of the system in three dimensions, our model is not significantly different than those previously presented in the literature (e.g. Bouwman et al. 2000 and Robitaille et al. 2007), which aimed to reproduce only the SED of the system. A breakdown of our model's best fit SED into its individual system components shows that from the near to far IR it is disk dominated (Figure \ref{SED}). Our model has minimal contributions from envelope material, which is consistent with \cite{Robitaille}'s AB Aur model. Therefore, our model, along with previous models presented in the literature, demonstrates that a disk dominated model can plausibly reproduce many of AB Aur's observed features, particularly at near-IR wavelengths.

\section{Origin of the Spiral Arms: Disk versus Envelope}
In Section 3 we showed that the envelope has little effect on the observed \textit{H}-band imagery and SED of the AB Aur system using Monte Carlo modeling techniques. In fact, Figure \ref{SED} shows that the envelope minimally contributes to the system's SED at mid- and far IR wavelengths, while the only significant contributor at \textit{H} band is the disk. In our models, the envelope density is low enough to not significantly affect the observed SED or morphology of the disk at \textit{H} band. Therefore, the only scenarios that are consistent with our modeling results are those that include an envelope with little to no material. This interpretation is also consistent with our low mass infall rate from the envelope ($10^{-7}$ M$_\sun$ yr$^{ -1}$), which is proportional to the envelope density and mass; we require a low mass infall rate, and therefore low envelope mass, to fit the observed data. These results are also in agreement with previous modeling attempts that can be found in the literature (e.g. Robitaille et al. 2007), which also find that a disk dominated model best reproduces many of AB Aur's observed features.

If such a low density envelope is the correct interpretation of the observed data, we believe that the disk, which has a much higher density and total mass than the envelope, is more likely to produce and maintain spiral structures, especially at AB Aur's current evolutionary stage. This supports the idea that at least some of the spirals are part of the disk.

Assuming that the spirals originate from perturbations within the disk, there exists several possible formation scenarios, such as disk-planet interactions, other gravitational instabilities, magneto-rotational instabilities, and accretion of material onto the outer regions of the disk. We consider the implications of the disk-planet origin by analyzing the behavior of multi-epoch \textit{H}-band scattered light imagery below.  

We obtained the previously reduced and published \textit{H}-band CIAO \citep{Fukagawa} and HiCIAO \citep{Hashimoto} imagery of the system to measure any potential rotation of the spiral features. Use of multi-epoch data in this way can constrain potential locations of planets within the system. By comparing the location of the spirals between the CIAO and HiCIAO data sets we are assuming that if there were existing polarimetric \textit{H}-band imagery from 2004, when the CIAO data were taken, that bright regions associated with the spirals would also show a larger polarization than the surrounding disk.

\begin{center}
\begin{figure*}[]
%\figurenum{4}
%\epsscale{0.75}
\centering
\includegraphics[width=\textwidth]{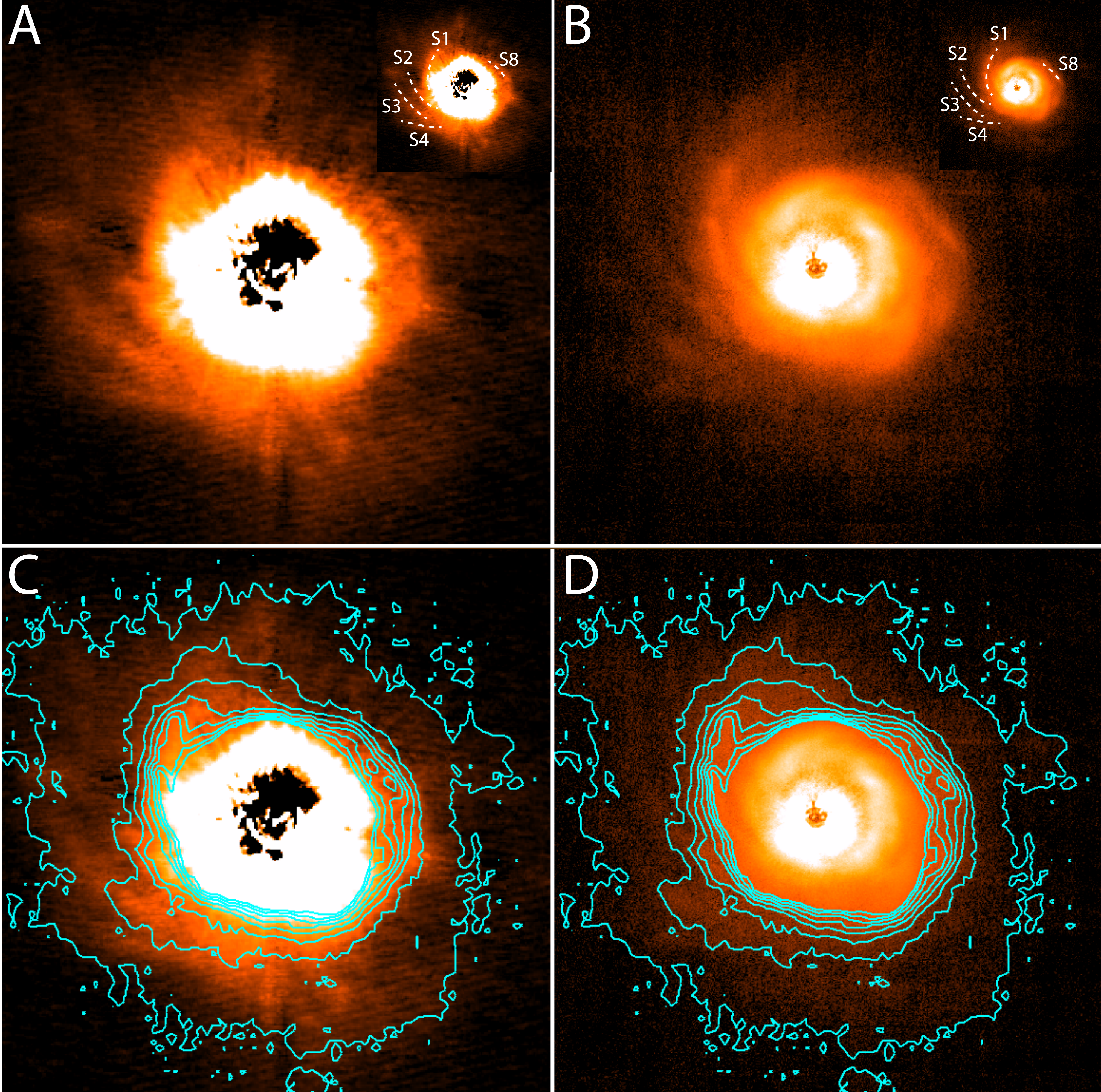}
\caption{Spiral structures in the \textit{H}-band observed imagery of the AB Aur system in total light (panel A; 2004 CIAO data) and polarized intensity (panel B; 2009 HiCIAO data). Insets in the upper right corner of panels A and B identify spiral structures using the same labeling scheme as Hashimoto et al. (2011). Panel C displays the 2004 CIAO imagery with a contour overlay of the 2009 HiCIAO data. Panel D shows the 2009 HiCIAO data with the same contour overlay as panel C. Comparision of the two \textit{H}-band images in panel C shows no significant movement of any of the marked spiral structures between the 2004 and 2009 epochs.} \label{SpiralMove}
\end{figure*}
\end{center}

Figure \ref{SpiralMove} displays the total light \textit{H}-band imagery from the CIAO instrument (panel A) taken on 2004 January 8 and 11 \citep{Fukagawa} and the \textit{H} band PI imagery from the HiCIAO instrument (panel B) taken on 2009 October 31 \citep{Hashimoto}. Several spirals are easily identifiable in both data sets. Therefore, we over plot the 2009 HiCIAO data on the 2004 CIAO imagery as contours in panel C, while panel D shows the HiCIAO contours over plotted on the HiCIAO imagery for comparison. Panel C shows no significant movement of any of the visible spirals over the 5.8 year baseline between our data sets.  

\begin{center}
\begin{figure*}[]
%\figurenum{4}
%\epsscale{0.75}
\centering
\includegraphics[width=\textwidth]{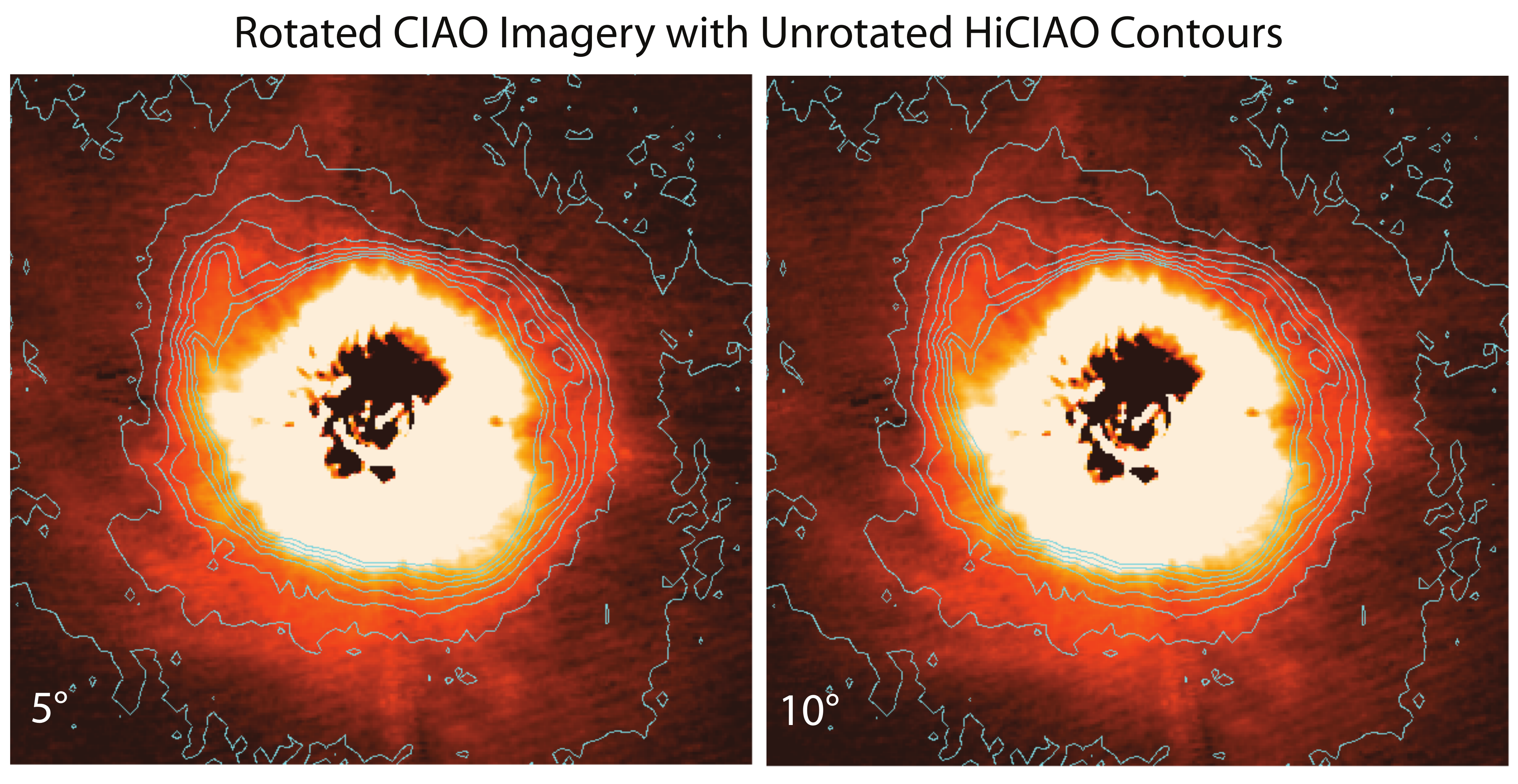}
\caption{\textit{H}-band CIAO total light imagery of the AB Aur system rotated counterclockwise by 5$\degree$ and 10$\degree$ about the central star with contours derived from the unrotated HiCIAO data overplotted. A lack of significant movement in the S1, S3, and S8 spirals between the two datasets suggests that if a planet is the cause of any of these structures, it must be at least 47 AU from the central star.} \label{Rot}
\end{figure*}
\end{center}

In order to better constrain this apparent lack of spiral arm movement, we deliberately rotated our imagery in Figure \ref{Rot} to determine what amount of rotation would have been detectable over our 5.8 year baseline. To do this, we rotated our CIAO images in the counterclockwise direction in $5\degree$ increments and overplotted the unrotated contours derived from the HiCIAO dataset. When comparing the rotated CIAO imagery to the unrotated HiCIAO contours, the effects of the rotation can be seen relatively quickly; it only takes $5\degree$ of rotation before the S3 and S8 contours do not line up with their respective spiral structures. At $10\degree$ the S1, S3 and S8 spirals all start to appear to be poorly fit, which only becomes more pronounced as the amount of rotation increases. However, a comparison of rotated HiCIAO imagery to unrotated HiCIAO contours (not shown in Figure \ref{Rot}) indicates that using datasets taken with similar observing techniques might allow us to detect a smaller amount of rotation; it only takes $5\degree$ of rotation before the S1 and S8 contours do not line up with the spiral structures. Needing larger rotations before the CIAO data appear to not be well fit by the unrotated HiCIAO contours is likely due to several effects. First, the pixel scale of the CIAO data is significantly larger than the HiCIAO data (21.3 mas and 9.3 mas respectively). Therefore, it would take a larger movement of material before it is noticeable in the CIAO data. Also, there are slight differences between the datasets which may hinder the detection of rotation to some effect; the HiCIAO imagery is polarized intensity while the CIAO imagery is unpolarized.

Regardless of these complexities, this analysis suggests that the S1 and S3 spiral structures have likely moved less than $10\degree$ over the 5.8 year baseline because we would have otherwise detected this amount of rotation. If the spiral structures are formed due the disk-planet interactions their motion should be in sub-Keplarian motion with a pattern speed given by $$\omega=\frac{360\degree}{2\pi}\sqrt{\frac{\textrm{GM}_\star}{\textrm{r}_\textrm{p}^3}},  \qquad \textrm{[deg\space yr}^{ -1}\textrm{]} $$ where $M_\star$ is the mass of the central star, $r_p$ is the orbital radius of the planet whose perturbations would invoke the spiral structure, and $G$ is the gravitational constant. Therefore, the orbital radius at which a planet might be located is given by $$\textrm{r}_\textrm{p}=\sqrt[3]{\left( \frac{360\degree}{2\pi}\right)^2\frac{\textrm{GM}_\star}{\omega^2}}=\sqrt[3]{\frac{3.107\times 10^5}{\omega^2}}. \qquad \textrm{[AU]}$$ This suggests that if planets are the cause of the S1, S3, and S8 spirals structures, they must be in an orbit located at least 47 AU from the central star. This  constraint is consistent with the 80 AU location of the gap which may have formed due to a planet clearing out disk material \citep{Hashimoto}.

If the disk-planet formation mechanism is correct and the planet were located in the middle of the gap at 80 AU, then it would take an approximately 6.4 year baseline between epochs to see a $5\degree$ rotation of the spirals (12.8 years for $10\degree$). We expect that future observations of the AB Aur system using techniques similar to existing data, e.g. a second epoch of \textit{H} band PI imagery, will provide more rigorous and clear constraints on the movement, or lack thereof, of the spiral arm structures. Recent modeling work by \cite{Dong} shows that a 3 $\textrm{M}_\textrm{J}$ planet at approximately 67 AU can excite spirals and form other structures that are very similar to those observed in AB Aur. At that distance the spirals should be moving at just over $1\degree$ per year. A new epoch of HiCIAO data taken as soon as this year should be able to determine if a planet at 67 AU forms the spiral structures; they should have moved by at least $6.3\degree$ now.

While beyond the scope of this paper, we feel that modeling of AB Aur's spirals using similar techniques to those used by \cite{Muto} for the SAO 206462 disk will have important implications for our understanding of the disk structure. In their work, \cite{Muto} used spiral density wave theory to predict how the future movement of the spiral patterns in SAO 206462 would deviate from the local Keplarian speed of the disk. In combination with future epochs of imagery, modeling such as this may further constrain or limit the potential formation scenarios for the spirals in the AB Aur disk. 

\section{Summary of Conclusions} 

We have used the \texttt{HOCHUNK3D} Monte Carlo code to self-consistently model the near-IR imagery and SED of the AB Aur system simultaneously. Our modeling results are consistent with those already present in the literature; a disk-dominated model reproduces many of AB Aur's observed features. This suggests that a spiral formation scenario involving disk material remains a possibility for at least some of the spirals, particularly in the \textit{H} band where the envelope does not significantly contribute to our model's SED or imagery.

Given our findings, we analyzed the 2004 and 2009 total light and PI imagery from the CIAO and HiCIAO instruments in the \textit{H} band to measure any potential rotation of the disk's spirals. In the event that the spiral structures are formed due to disk-planet interactions, the spirals' movement, or lack thereof, can constrain the locations of possible planets within the AB Aur system. We find no significant rotation of any of the spiral structures over the 5.8 year baseline between these two datasets. By purposely rotating our data and comparing them to unrotated versions of the data, we find that if the spirals did move, they did so by less than $10\degree$. This suggests that if a planet were responsible for the observed structures, it is in an orbit that is least 47 AU away from the central star.

\acknowledgments

We acknowledge support from NSF-AST 1009203 (J.C.), 1008440 (C.G.), and 1009314 (E.R, J.W, J.H) and the NASA Origins of Solar System program under NNX13AK17G (J.W.), RTOP 12-OSS12-0045 (M.M.),  NNG16PX39P (C.G), and NNG13PB64P (C.G.). This work is partly supported by a Grant-in-Aid for Science Research in a Priority Area from MEXT Japan and by the Mitsubishi Foundation. The authors recognize and acknowledge the significant cultural role and reverence that the summit of Mauna Kea has always had within the indigenous Hawaiian community. We are most fortunate to have the opportunity to conduct observations from this mountain. We wish to extend special thanks to those of Hawaiian ancestry on whose sacred mountain we are privileged to be guests. This work is based in part on data collected at Subaru Telescope, which is operated by the National Astronomical Observatory of Japan. We also thank Barbara Whitney for providing us with helpful feedback that improved our paper and clarifying aspects of her HOCHUNK3D code and Anthony Paat for helping run models. Additionally, we would like to thank the anonymous reviewer for providing comments that led to an improved paper.

\end{document}